# ANALYZE BUSINESS CONTEXT DATA IN DEVELOPING ECONOMIES USING QUANTUM COMPUTING

*Author: Ammar Jamshed*

*MSc Data Science Student at University of London, Goldsmiths*

## Abstract

Quantum computing is an advancing area of computing sciences and provides a new base of development for many futuristic technologies discussions on how it can help developing economies will further help developed economies in technology transfer and economic development initiatives related to Research and development within developing countries thus providing a new means of foreign direct investment(FDI) and business innovation for the majority of the globe that lacks infrastructure economic resources required for growth in the technology landscape and cyberinfrastructure for growth in computing applications. Discussion of which areas of support quantum computing can help will further assist developing economies in implementing it for growth opportunities for local systems and businesses.

## Table of Contents

# Table of Figures

# Introduction

Quantum computing as a discipline started in North America and was used to crunch Large scale into analyzable segments of trends such as data of space anomalies like wormholes as the length and storage of so much data require very powerful machines and large-scale data storage systems for which quantum computing came to be used as it could crunch down data into quantum bytes usable for machine and algorithms to the computer which otherwise were considered humanly impossible to view and understand

Quantum data is created with the use of quantum circuits which are layers of circuits within traditional circuits that process the data into qubits (Quantum Bits).

\

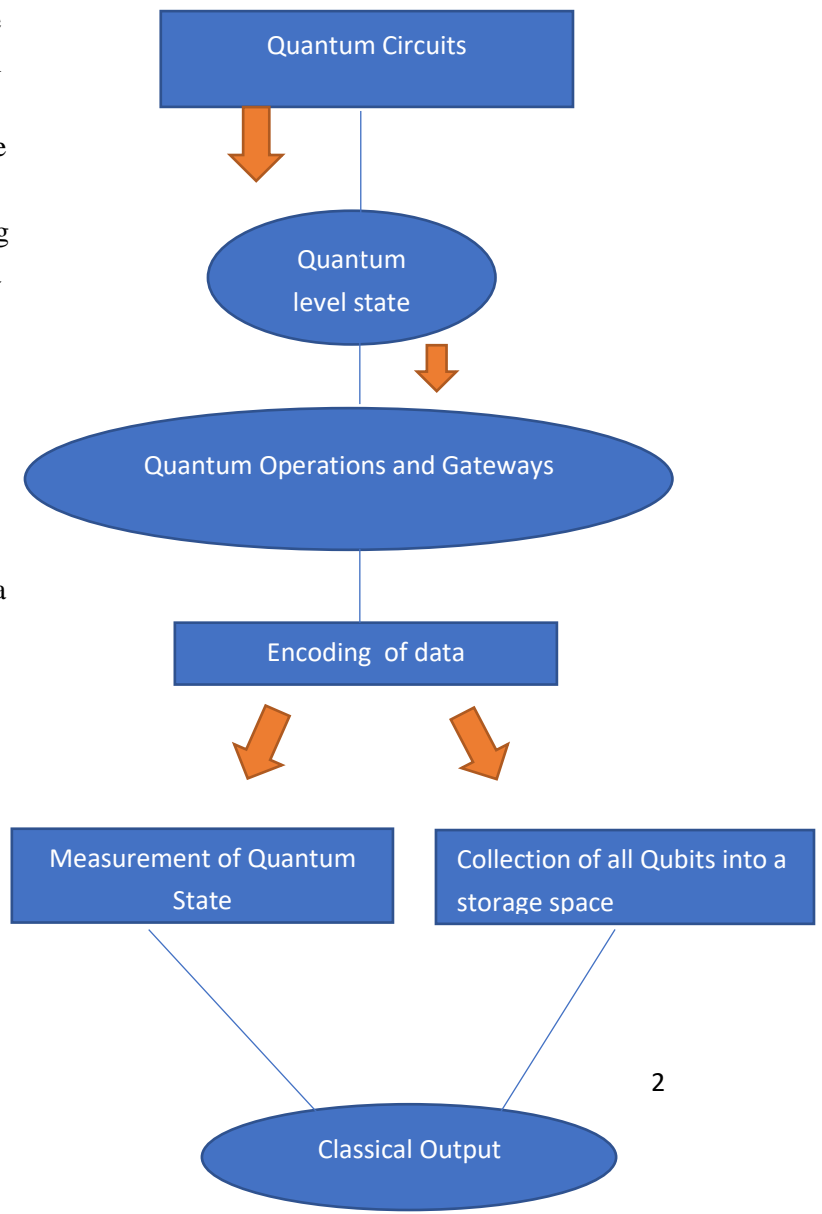

Figure 1: Quantum Data Generation

# How Quantum Computing has overtime expanded into Business frameworks

Quantum computing initially started for large-scale machine programing and analyzing high-grade weaponry for wars It over time expanded into space research and usability in large-scale digital storage management for public institutions and social networks. Quantum computing has applications in advanced computation of business data analytics and predictions along with estimations of event likelihoods that could impact the business.

### 1. Simulation of Scenarios

Using simulation for designing new drugs and material compositions for new chemicals or test pilot test experimental outcomes before conducting them in laboratories.

### 2. Optimization of Processes

Logistics of transport management for vehicles across countries, scheduling of millions of automated tasks for Billions of transactions worth of data and managing national portfolios and Countrywide portfolios of investments and finances can be done with Quantum computing platforms.

### 3. Predictability of volatile events

Events that are too volatile and otherwise cannot be predicted like stock market fluctuations from uncertainty-based events or natural calamities due to changes in earth atmospheric variables can be predicted through quantum data computations by micro scaling variables and their relationships.

### 4. Computing Storage Feasibility

With the use of quantum circuits, we can create vast storage spaces that can store 0 and 1 values of data at the same time due to the superposition effect of quantum computing.

### 5. Computing infrastructure

We previously relied on Moore's law of computing power to predict data hardware changes to storage space requirements in terms of data processing power doubling every 18 months and Metcalfe's law of law on communication networks will double after every user addition to the network chain. Quantum computing may make those estimators redundant and prove exponential growth rates in networks and hardware optimization beyond basic statistical estimates on computing.

Figure 2: Process Relationship between the quantum computer and volatile Data

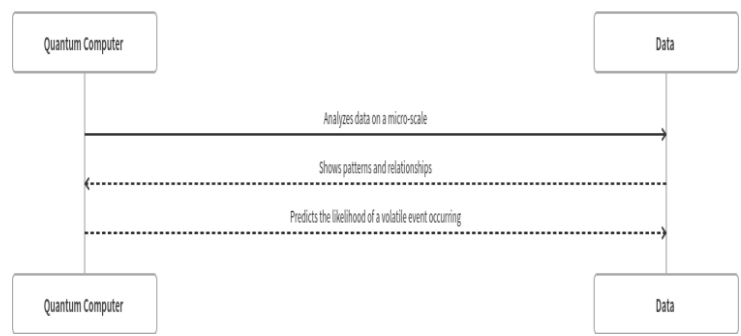

# Challenges of Quantum data is being used in business Applications within Developing economies

Quantum computing is very expensive and requires large-scale power usage and maintenance as it heats up very fast and requires extremely cold temperatures to operate and prevent machines from overheating thus beyond costs its technical management expertise that may become a barrier for it to be used in developing economies.

Skills and training for quantum computing are very limited due to its recent emergence as a field and as the infrastructure for it is not readily available thus making it hard to incubate training centers for its general use. Universities are offering more degrees and modules related to quantum science-based training, but they mostly dwell on theoretical research at a global scale rather than practical implementation.

Quantum computing requires high-powered electricity and low-temperature environments to operate which are not only expensive to set up and run but require adhering to many regulatory guidelines of usage. Superconducting quantum computers like those constructed by giant tech conglomerates like Google and IBM require helium-3 to run and operate which is a nuclear by-product.

Quantum computing offered on cloud services and remote solutions too aren't available in full capacity for businesses and can't be used solely to build high-grade quantum solutions for effective business management. Furthermore, issues of scalability exist for quantum cloud services in terms of data migration and workflow training.

# Quantum computing works in developing economies

While Quantum science research and applications are often stereotyped as only capable of being conducted in highly developed economies developing economies have also contributed significantly to the development and usage of its benefits in science and technology as well as research applications and a few of them within regional importance are listed below.

### i. Pakistan

Pakistan has been working on Supercomputing solutions and uses since project 706 of its nuclear program since the 1970s.

In 2012 National University of Science and Technology developed the fastest supercomputing facility in Pakistan which runs to date thus paving the way for quantum computing to initiate as well.

In 2016 Pakistan Atomic Energy Commission announced its initiative to work on a quantum computer to use for secure nuclear communications.

Pakistan's Top Business school LUMS hosted an international conference on quantum computing with world-renowned experts for a panel discussion.

In 2020 National University of Science and Technology in Islamabad launched the Center for Quantum Research and Development (CQR&D) to work on developing quantum technologies.

### ii. Iran

Iran has made advancements over the years in quantum computing as well as nanotechnology despite being under sanctions. In 2021 Iran opened the National Center for Quantum

Technology to promote ongoing efforts in the laboratory studies and industrial applications of quantum computing within Iran.

Iran had previously made claims of building a new quantum computer in 2019 which was later retracted after an investigation found out it was only a classical computer programmed with quantum operations which was revealed by multiple media sources.

In 2020 Dr Javad Karimi Sabet who is AEOI deputy and head of the Nuclear Science and Technology Research Institute in Iran claimed in a public press that Iran is making progress in application of quantum computing applications which include quantum coding in Fiber ground, quantum navigation, metrology, radar activity simulation along with the atomic clock and quantum biology.

### iii. Malaysia

Malaysia is one Eastern country that is beginning to make its mark in quantum computing, Malaysian government and private sector are making strides with quantum technology in Malaysia. In December 2022 the International Conference on Quantum Engineering and Technology took place in Kuala Lumpur, Malaysia as the scientific climate in Malaysia is proving to manifest for quantum technology researchers and Investors.

Universiti Malaya in Malaysia has a dedicated department unit for studying Quantum laser and quantum computing applications.

### iv. Egypt

Egypt started the Alexandria Quantum Computing group within Alexandria University to encourage and develop more research in quantum computing from Egypt to help improve Egypt's focus on quantum computing as well as popularize it within other regions, which have developed multiple courses and publications in Quantum computing. It launched QEgypt later in April 2021[c].

There have been multiple startup businesses opened in Egypt to provide quantum computing services as market forecasts are seeing the potential of quantum computing business growth in Egypt.

Egypt joined the quantum flagship program in collaboration with the EU to contribute to the Quantum science community After its launch in 2018.

### v. Rwanda

Rwanda has worked on promoting science and technology as part of its national policy and objectives over the years since 2015.

Rwandan government in Partnership with the African Institute of Mathematical Sciences in 2019 launched Quantum Leap Africa to promote African scientists' work in quantum computing which has trained many African students over the years in Quantum computing.

In 2022 during the World Telecommunication Development Conference in Kigali, Rwanda. The government and many of its representatives spoke about initiating quantum computing labs in Rwanda to build an effective quantum computing infrastructure in Rwanda.

### vi. Ghana

Ever since becoming part of OPEC, Ghana has made significant improvements in technologies and infrastructure.

There are multiple Universities in Ghana not limited to the African Institute of Mathematical Sciences Ghana that have drafted master's degrees and other training programs in Quantum computing or quantum science-based disciplines.

While Ghana is still in its early stages compared to its counterparts it is still progressing towards improvements in applications of Quantum computing and will over time prove to be well acquainted with state-of-the-art hardware for quantum research.

# Literature review

I. According to the analysis by Bhasin, A., & Tripathi, M. (2023), There are various technologies deployed with the understanding of quantum computing principles even advanced financial models by research accumulations of Turkish researchers there has been an analysis on the management aspect of technologies inclusive of quantum computing and how data governance is required for its management and deployment in real world settings and many projects done on its regards have not given a feasible return due to the limited understanding of Quantum computing usability along with the risks it carries not limited to running costs.

II. The need for quantum computing technologies and training for workforces in it has been discussed in detail by Peterssen, G. (2020) in terms of the skills and training required in educational settings and applied practices of quantum computing requirements and the data analysis of his research showed that while quantum computing is still restricted to developed economies there is still transfer happening and developing economies and Asia/Pacific region is showing high rates of interest and research within quantum computing with the leaders being China, India, South Korea and Japan but there is still need for it to transfer to developing economies to function the global economy values chains as developing economies with highly dense populations are the major markets for western businesses and require technology transfers at all length to ensure a proposers future.

III. Merging Quantum computing with blockchain ledger would create a powerful model for distributed computing and an effective way for data processing to further optimize and improve as discussed by Mosteanu, N.R.; Faccia, (2021) in their research publication.

IV. Research and analysis by de Wolf, R. (2017) have shown that cryptography, and system simulation optimizations of Quantum systems are the key areas of how it will transform, societal mechanisms that would alter the society paradigms of networking and value chains thus from the biggest establishments to the smallest exchanges would involve the use and security of quantum computing within society shortly based on the current forecasting trend of quantum uses. It also evaluates the risks of data usage that will erupt with the transformation cycle towards quantum computing optimization especially for developing economies as their cyber infrastructure and guidelines aren't as applicable as the developed economies.

V. Taking Inspiration from a Quantum computing composite from a single neuron, Research by Zidan, M., Abdel-Aty, A., El-Sadek, A., Zanaty, E. A., & Abdel-Aty, M. (2017) has shown low-cost output in quantum computing can be achieved which will make it more economically feasible for small businesses and developing economies as well in terms of applying predictive modelling and systematic changes with quantum computing.

VI. The analysis of Gallium Nitride(GaN) as raw input and its potential effects on technological advancements in e3lectrinics and smart devices as per the research study by Teo, K. H., Zhang, Y., Chowdhury, N., Rakheja, S., Ma, R., Xie, Q., Yagyu, E., Yamanaka, K., Li, K., & Palacios, T. (2021), would help technology acceleration in consumable uses of quantum computing in common device platforms.

VII. A Research study by Si, Y., Wang, R., Zhang, S., Zhou, W., Lin, A., & Zeng, G. (2022) has examined that by using quantum computing to optimize energy management processes in maritime activities involving deep sea travel and drilling would help in developments of new maritime infrastructure and economic zones that may have been inaccessible and would boost global trade as it helps develop hybrid energy systems including those reliant on fossil fuels which majority trade by sea shipments today.

VIII. The usability of Linear Quantum computing is analyzed in the research by Giani, A., & Eldredge, Z. (2021), It can be examined that Faults do lie in the infrastructure design built also examined by the necessity of a global value chain to procure components and services to set up for large scale industrial projects. The structure and design of Linear Quantum computing may try to minimize the cost function within any predictive modelling framework. The design for electricity flow running Qubits in a parallel computing state or within a Quantum superstate may create room for major errors without a back door loop within the bell circuit stage.

IX. Application of quantum computing can create many commercially viable solutions as analyzed by Cusumano, M. A. (2018), a pair of qubits can represent four different states simultaneously and while we may lack the capacity in current storage levels in computing devices manufactured with storage limitations becoming a wide issue for higher scale R&D the implementation of Quantum data center's for backing up commercial services may not be very far.

X. According to research by Bova, F., Goldfarb, A., & Melko, R. G. (2021), Quantum computing may not be available at the scale required for use cases in banking and advanced manufacturing but a few areas where it's now being initialized within is quantum-safe encryption and drug discovery and overtime with cost optimization in supply chain ecosystem for hardware manufacturing quantum comping may become usable to address the problem of 'Combinatorics' which sets to measure the infinite number of ways the objects can be combined.

XI. According to Jenkins, J., Berente, N., & Angst, C. (2022, January 4), firms are beginning to adopt strategies towards implementing quantum computing for business operations despite quantum computing still being available at full scalability required for it and those strategies are characterized as option, discovery and adversarial strategies depending upon the business stage of operations.

XII. Research by Bayerstadler, A., Becquin, G., Binder, J., Botter, T., Ehm, H., Ehmer, T., Erdmann, M., Gaus, N., Harbach, P. H. P., Hess, M., Klepsch, J., Leib, M., Luber, S., Luckow, A., Mansky, M., Mauerer, W., Neukart, F., Niedermeier, C., Palackal, L., . . . Winter, F. (2021) on Industrial usage of quantum computing have shown analysis on efforts by the German government in establishing Quantum Technology and Application Consortium (QUTAC) in a public-private partnership with industrial leaders to decide on technology

XII. components and systemic for quantum computing to be used and implemented by businesses.

XIII. The research paper by Raheman, F. (2022) highlights the computing architecture and systematic structuring around quantum computing that show vulnerabilities around third-party applications carrying cyber threats and it comes up with a solution of Zero-vulnerability computing (ZVC) to combat the situation and banning third party application by building an effective solid-state software on a chip allowing a machine to handle any quantum threats or malware.

XIV. Findings shown by S. Gupta and V. Sharma at the 4th International Conference on Intelligent Engineering and Management (2023) analyzed the context of how quantum computing is becoming impertinent to many sectors of business and industrial use cases alongside challenges that mainly comprise public policy and cost management surrounding the implementation of Quantum computing applications and their availability.

XV. Aljaafari, M. M. (2023) analyzed social business models that can be improved with the use of quantum computing to analyze large chunks of data to develop effective tools for computational business management.

XVI. Research by Mohseni, M., Read, P., Neven, H., Boixo, S., Denchev, V. S., Babbush, R., Fowler, A. G., Smelyanskiy, V., & Martinis, J. M. (2017) five years ago claimed that three positions in the use of quantum computing will emerge in commercial computing that would Quantum Sampling, Quantum Assisted optimization and Quantum simulation. While some of the expected outcomes are materialized in scenarios today such as being available in cloud computing resources for simulation and quantum-assisted optimization along with Quantum sampling, they have not become fully available for commercial use cases in 2023 and are still restrictive in deployment capabilities due to common hardware limitations of Quantum states.

XVII. Awan, U., Hannola, L., Tandon, A., Goyal, R. K., & Dhir, A. (2022) developed a new approach employing the fuzzy analytical process of hierarchy to analyze challenges within the adoption of quantum computing across industries and discovered that lack of technical expertise and

XVIII. Nagori, V., & Varadarajan, V. (2023) researched possibilities for quantum cryptography to be in use when quantum technologies become widely available and how businesses must manage their data security with quantum cryptography as a cyber computing defense against malicious attacks.

XIX. Chauhan, V., Negi, S., Jain, D., Singh, P., Sagar, A. K., & Sharma, A. K. (2022) review quantum computing architecture and analysis implementation use cases for it in different industries and their research provides insights into both benefits and challenges of its implementation while algorithms in quantum computing such as Shor's Algorithm and Grover's Algorithm have multiple use cases even beyond the depths of quantum computing as AI deployment with quantum computing will enhance scalability of global systems in various aspect with these algorithms employed for its analytical capabilities.

accuracy are the main concerns in its adoption by organizations whose interest are aligned around these surrounding variables within their business processes.

# Future Implications of Quantum Computing in Business Applications

With time as technology progresses and the global economy becomes increasingly integrated, we are looking at times when quantum computing power would be available in all households by 2050 or so. Multiple theories and views by experts are argumentative on the use of quantum computing on whether it will not work in favor of society mechanisms and can further advance cyber-attacks are more massive scales[xviii].

We can look at advancements of many existing and innovative technologies as a result of more usability of quantum computing applications such that we can see more secure cyber security platforms preventing overloads of login and attempts to hack using digitally injective or analogue mechanisms as Quantum computing technology requires a lot of computing power to load and also a lot of time to allow any form unauthorized access or program to be injected into its scripted functioning thus it would take another quantum computing powered gateway to be able to hack into another quantum machine and such things are very low in supply and are procured at high costs thus leaving a lot of traceability for Any such hacking to take place.

Machine learning and deep learning models will become more powerful in computing predictive trends and data analytics as any form of data would be understandable for the program with no storage limitations with quantum power core machines[xix].

Developing economies with technology transfer, training and development of workforces will eventually establish proper data governance and computing guidelines which will help quantum computing applications merge into the developing economy framework in terms of business management as well as technological administration.

# Appendix

1. **Media Sources for cross-referencing Quantum computing work in developing economies.**

    a) Library (2023) Iran Watch.
    b) Organization (2021) Iran opens National Center for Quantum Technology, Tehran Times.
    c) Qegypt (2021) Qworld.
    d) World Telecommunication Development Conference 2022 (WTDC-22) - ITU (2022).
    e) AIMS Ghana.
    f) The Quantum Revolution (2021) SBASSE.